\documentclass[pre,aps,showpacs,twocolumn,floatfix,superscriptaddress]{revtex4}
\usepackage{epsfig}
\usepackage{color}
\usepackage{amsmath,amssymb}

\newcommand{\be}{\begin{equation}}
\newcommand{\ee}{\end{equation}}
\newcommand{\bea}{\begin{eqnarray}}
\newcommand{\eea}{\end{eqnarray}}
\newcommand{\ba}{\begin{array}}
\newcommand{\ea}{\end{array}}

\begin{document}
\title{Dynamical model for the full stretching curve of DNA}

\author{Alessandro Fiasconaro}
\email{afiascon@unizar.es}
\affiliation{Departamento de F\'{\i}sica de la Materia Condensada,  Universidad de Zaragoza, 50009 Zaragoza,  Spain}
\affiliation{Instituto de Ciencia de Materiales de Arag\'on,  CSIC--Universidad de Zaragoza, 50009 Zaragoza, Spain}

\author{Fernando Falo}
\affiliation{Departamento de F\'{\i}sica de la  Materia Condensada, Universidad de Zaragoza, 50009 Zaragoza, Spain}
\affiliation{Instituto de Biocomputaci\'on y F\'{\i}sica de Sistemas  Complejos, Universidad de Zaragoza, Zaragoza, Spain}

\date{\today}
\begin{abstract}
We present a phenomenological dynamical model able to describe the stretching features of a length \textit{vs} applied force DNA curve. As concerning the chain, the model grounds on the discrete worm-like chain model with the elastic modifications, which properly describes the elongation features at low and intermediate forces. The dynamics is developed under a double well potential with a linear term, which, at high forces, accounts for the narrow transition present in the DNA elongation (overstretching). A quite good agreement between simulation and experiment is obtained.
\end{abstract}

\pacs{87.15.-v, 36.20.-r, 87.18.Tt, 83.10.Rs, 05.40.-a}
\keywords{Stochastic Modeling, Fluctuation phenomena, Polymer dynamics, Langevin equation}


\maketitle

{\it Introduction}. The stretching curve of DNA has been the subject of many studies, both theoretical and experimental. The first experimental study of the DNA stretching was performed by Bustamante and co-workers \cite{Busta1992}, where the DNA length was measured as a function of the force applied to the chain. From the theoretical point of view, the small and intermediate force range of the DNA stretching curve can be understood and depicted by means of the wormlike chain (WLC) model \cite{wlc}, where the DNA is modeled as a flexible beam. For simulation purposes, the WLC model can be discretized as a chain of beads connected by sticks with the presence of an elastic bending (discrete WLC) and improves the more naive freely joined chain model \cite{fjc}). This WLC model (and the discrete version) fits with excellent precision the experimental curve, making use of the correction introduced by Odijk \cite{odijk} that takes into account the extensibility of the chain when the sticks are replaced by harmonic springs.

The most intriguing feature of the stretching of DNA is presented
in a subsequent experiment of Bustamante and co-workers
\cite{Busta1996,Cluzel1996}, where a sudden elongation of the
chain is registered when a large force, around $70 pN$, is
applied.

The presence of this tension-induced \emph{overstretching
transition} reveals the existence of two structurally different
DNA states. The first state, at low applied forces, is the so
called B-DNA conformation, where the base pairs are helicoidally
packed in their native state. The other state, present at high
forces, was first called the S state, where the base pair
distance is higher by about $70\%$ than that in the state B.
The nature of the overstretched DNA state remains controversial,
and it seems that many effects could be responsible for the sudden
DNA elongation at large forces.

Molecular dynamics simulations revealed that the DNA transition
could correspond to a change of the helicoidal DNA shape to
another one with a shorter helix radius and a large tilt of the base
pairs \cite{Cluzel1996}. This result seems also be confirmed by
early spectroscopic studies \cite{fraser1951}.

On the other hand, the suggestion that the enlargement is
accompanied by a rotation of the bases giving rise to a ladderlike
structure in the S state of DNA, has also been made
\cite{konrad1996}. Motivated by the experimental data of Strick \textit{et
al.} \cite{strick1996}, a possible linear coupling of the
twist-stretch variable was proposed \cite{kamien1997, moroz1997},
and in this last paper an estimation of the DNA twist stiffness
was also performed. More accurate measurements suggest that a
not completely ladderlike form of the DNA occurs in the
elongation from the B- to the overstretched DNA state. In fact a
remanent helicity of the DNA persists in its overstretched state
of about one turn every 37.5 base pairs, almost four times smaller
than the natural helicity of one turn per 10.5 base pairs
\cite{leger1999}. Moreover, by avoiding the writhing of the DNA
double helix, a broadening in the overstretching transition is
observed.

Various statistical-mechanics models have been implemented in
order to mathematically describe the narrow overstretching
transition, which take into account the WLC model
\cite{Cizeau1997,leger1999,storm2003}, with an increasing number
of parameters in order to take into account the twist constraints
\cite{moroz1997,marko1998}.

The existence of the S-DNA state as a hybridized state of DNA has
been the object of deep and still controversial investigation, mainly
arising because of the presence of an asymmetric hysteresis in
the pulling and relaxing DNA elongation curve, which appeared to
indicate a force-induced DNA melting
\cite{MetzlerPRL2008,WilliamsBJ2001,WennerBJ2002}. Further
investigations, both numerical and experimental, were published up until very recently, some of which confirmed
\cite{WilliamsPNAS2009,vanMameremPNAS2009,CoccoPRE2004,MarenduzzoPRE2010},
and some did not confirm \cite{FuNAR2010,BiancoBJ2011,ZhangPNAS2012} this hypothesis.

\begin{figure}[h]
\centering
\includegraphics[width=8.2cm]{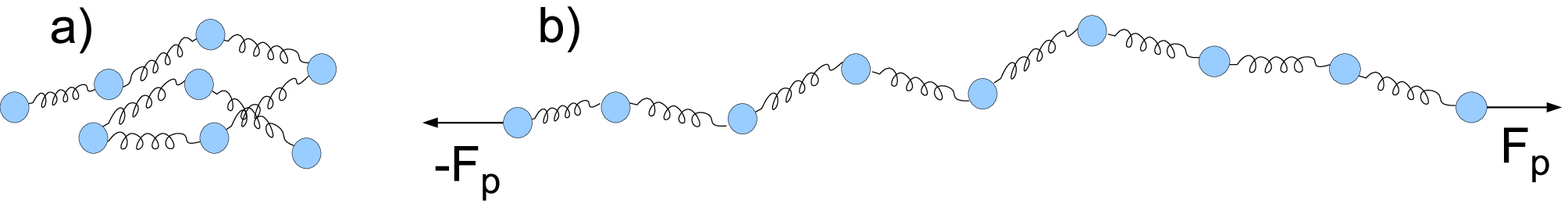}
\caption{Scheme of linear chain in 3D space. In (a) the chain is free to move in the thermal bath, while in (b) is pulled from each end with a force $F_p$.}
 \label{chain}
\end{figure}
In this work, we present a dynamical Langevin model where a polymer is pulled under an asymmetric free energy potential, with the purpose of describing the narrow DNA overstretched transition at high forces in a Landau-Ginzburg landscape. This model proposes a simple phenomenological way to describe the DNA dynamics, and its results are in good agreement with the most relevant experimental outcomes. The model we present doesn't enter in the microscopic details of the DNA elongated state, if melting is or is not responsible for it. The description presented is valid regardless of the specific microscopic state of the DNA. In this sense the model is general and can be included as an ingredient into other more complex processes like DNA translocation driven by molecular and nanotechnological motors.

\begin{figure}[tp]
\centering
\includegraphics[angle =-90,width=8.2cm]{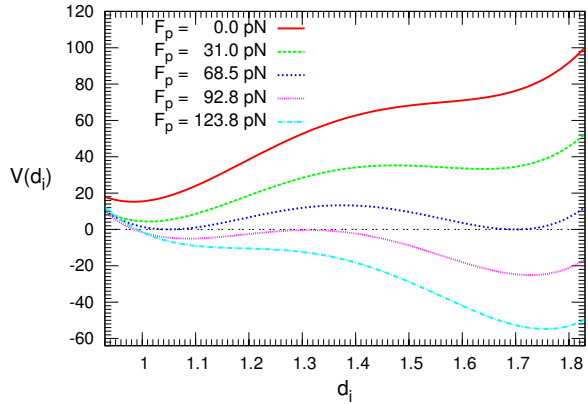}
\caption{Potential for different values of the force $F_p$ acting on the chain. The forces increase from the upper to the lower curves.}
 \label{pot}
\end{figure}

\emph{The chain model. }We model the DNA molecule by a three-dimensional (3D) polymeric chain of $N$
dimensionless monomers connected by nonharmonic springs (\ref{chain}). In spite of the usual harmonic interaction \cite{Rouse}, our elastic potential energy is given by
 \be
 V_{\rm el}(d_i)=\frac{k_e}{2}\sum_{i=1}^{N} (d_i-l_0)^2 (d_i-l_1)^2  + k_l d_i,
 \label{v-har}
 \ee
\noindent where $k_e$ is the elastic parameter, $\mathbf{r}_i$ is the position of the $i$-th particle, $d_i = |\mathbf{d}_i| = |\mathbf{r}_{i+1}-\mathbf{r}_i|$ is the distance between the monomers $i+1$ and $i$, and $l_0$ and $l_1$ are minima of the potential representing approximately the equilibrium distance between adjacent monomers, at weak and strong forces respectively. The potential is plotted in Fig.~\ref{pot}. The parameter $k_l$ is chosen in such a way that the middle of the force transition corresponds to equal probabilities to cross the elongation potential barrier from either \emph{left to right} (enlargement) or \emph{right to left} (constriction).

The model takes into account the bending energy with a term given by
 \be
 V_{\rm ben}(\theta_i)=\frac{k_b}{2}\sum_{i=1}^{N} [1- cos(\theta_i-\theta_0)],
 \label{v-ben}
 \ee
where $k_b$ is the bending elastic constant, and $\theta_i$ is the angle between the link $\mathbf{d}_{i+1}$ and the link $\mathbf{d}_{i}$.

The dynamics of the chain is given by the overdamped equation of motion
 \be
 \dot{\mathbf{r}_i} = - \mathbf{\nabla}_i V_{\rm el}(d_i) -\mathbf{\nabla}_i V_{\rm ben}(\theta_i) - \sqrt{2k_BT} \xi(t),
 \label{eq}
 \ee
where $\xi(t)$ represents the thermal contribution as a Gaussian uncorrelated noise, and the time $t$ is scaled with the damping $\gamma$ as $t \rightarrow t/\gamma$. Time units are then given in $s \gamma^{-1}$. The nabla operator is defined as $\mathbf{\nabla}_i = \partial / \partial x_i \mathbf{i} + \partial / \partial y_i \mathbf{j}  + \partial / \partial z_i \mathbf{k}$.
Two forces $F_p$ and $-F_p$ act on the chain on the first and last monomer respectively in order to stretch the polymer. The figures show the length of the DNA polymer as a function of the applied force $F_p$.

\begin{figure}[tbp]
\centering
\includegraphics[angle=-90, width=8.1cm]{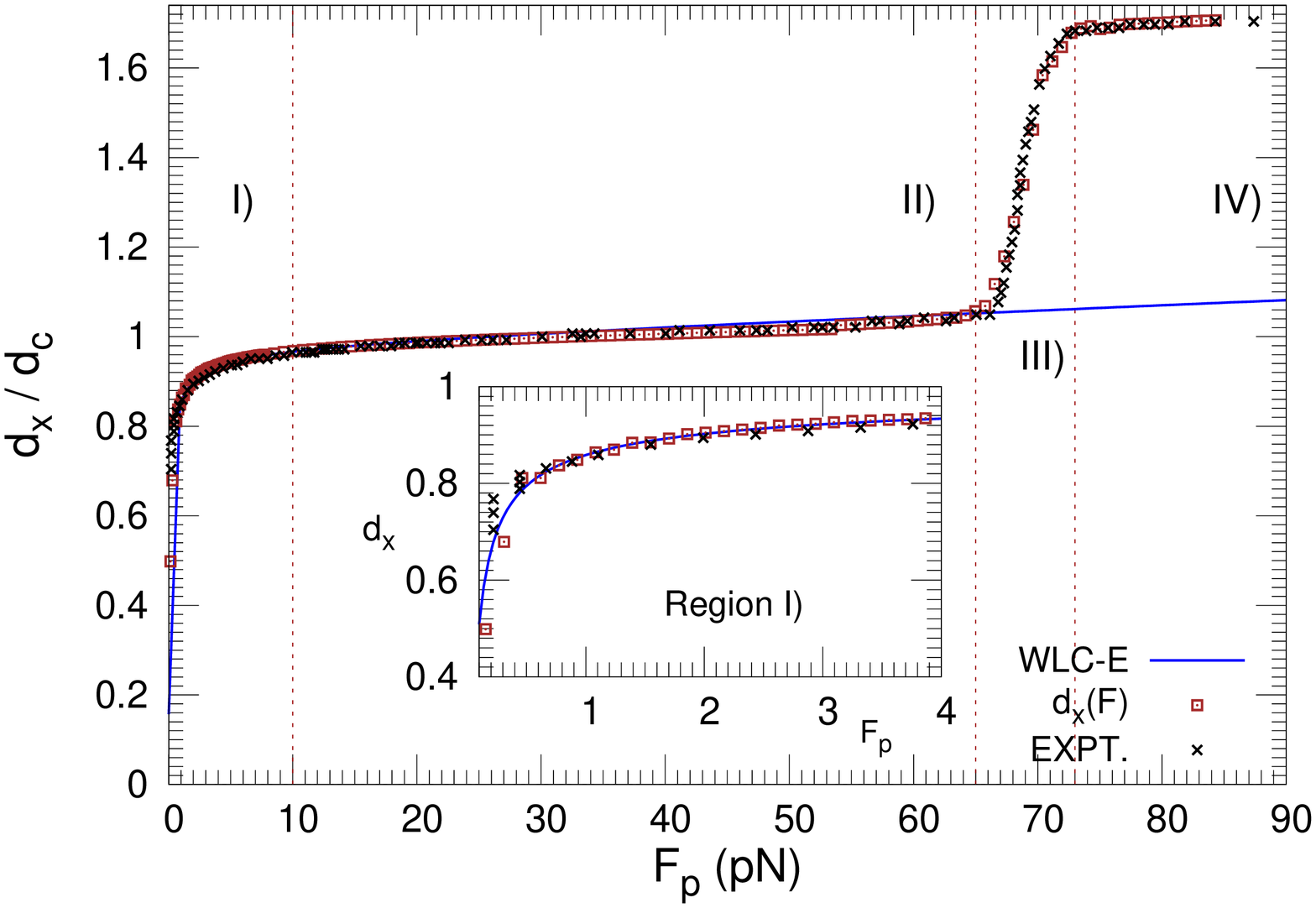}
\caption{DNA length vs Force. Parameters: $t_s = 5 \times 10^6$, $N=100$, $k_BT = 4.1$, $k_b=10$, $k_e=1190$, and $k_l=89$ (units given in the text).}
\label{DNA-all}
\end{figure}
\emph{Results. }We performed various computer simulations with a
Runge-Kutta stochastic algorithm for a long fixed simulation time
$t_s= 5 \times 10^6$ time units (t.u.), and $dt=0.01$. We set the units of energy and length to be $E_u = 4.1$~pN~nm (thermal energy at room
temperature) and $l_u= 5.3$~nm, respectively. With these units the
parameters used in simulations are $k_BT= E_u$,
$k_e=1190E_u/l_u^4$, $l_0= l_u$, $l_1=1.7l_u$, $k_b=10 E_ul_u$,
and $\theta_0=0$. Thus the persistence length of the chain is $\xi
= k_b/k_BT = 10 l_u=53$nm, which gives the correct WLC model fit
from the experimental data. The parameter $k_e$ is determined from
the extended WLC model in order to fit the DNA elongation data in
region II, and analogously for the bending parameter $k_b$.

After a thermalization time of $t_s=1 \times 10^6$ t.u., the
average of the lengths at all subsequent times has been
evaluated with respect to the contour length $d_c$ of the polymer,
with $d_c = (N-1) l_0$, and the values are plotted in
Fig.~\ref{DNA-all}, where a very good agreement between the
computer simulation  (brown $\boxdot$) and the experiments
(black~$\times$) is shown \cite{Busta1996}.

\begin{figure}[b]
\centering
\includegraphics[angle =-90, width=8.2cm]{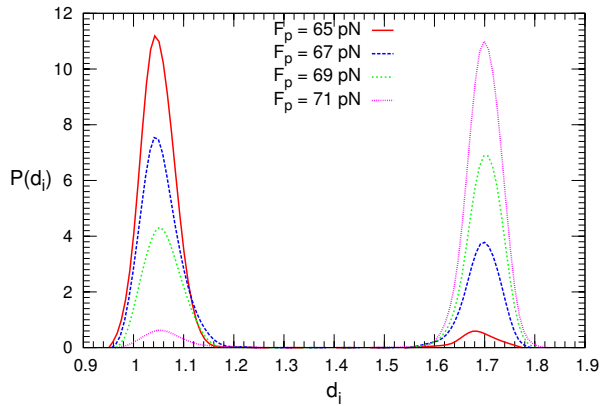}
\caption{Probability density of the length of the links $d_i$ for different applied forces along the overstretching transition.}
 \label{histo}
\end{figure}
The inset of the figure shows good agreement also at low force values. The full line represents the prediction of the WLC model in region I and II, which follows very well both the simulations points and the experimental ones.
Figure~\ref{histo} shows the density distribution of the link length close to the transition. It is there evident that a gradual increase of the number of enlarged links (\textit{i.e.}, $d_i \approx 1.7$) is produced by increasing the applied pull force $F_p$.

\begin{figure}[pt]
\centering
\includegraphics[angle =-90, width=8.2cm]{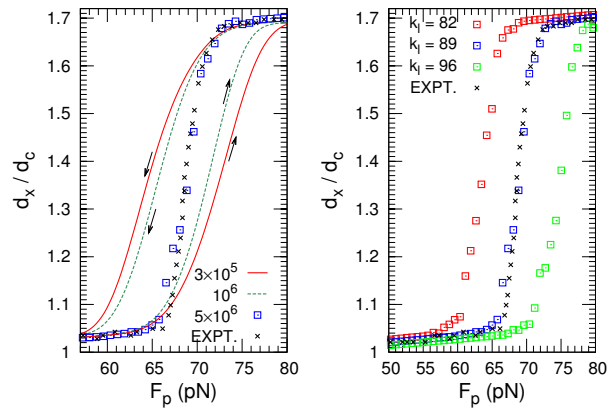}
\caption{Left panel: transition region with different application times of the force, $t_s = 3 \times 10^5$, and $t_s = 10^6$, shorter than the relaxation time of the chain. Right panel: the same for different values of the potential parameter $k_l$, $k_l$ increasing from left to right curves.}
 \label{ist}
\end{figure}

The transition between the regimes II (B-DNA) and IV (overstertched) arises in the very short force range.
We model this feature using the Ginzburg-Landau (GL) effective potential given in Eq.~(\ref{v-har}). This approach has also been used to describe the B-DNA to Z-DNA transition \cite{may2010}. In the GL phase transition theory, an asymmetrical contribution to the free energy gives rise to a discontinuous (first-order) phase transition with the temperature $T$ as control parameter \cite{chaikin}. Analogously, in the present case, we obtain a sharp transition with the DNA elongation $d_x$ as order parameter, and the force $F_p$ as control, with the difference with respect to the standard behavior that the intensity of thermal fluctuations is here an external parameter, whose presence causes a relatively smooth transition between the two states.

A dynamical fingerprint of discontinuous transitions is the appearance of hysteretic behavior due to a slow convergence to equilibrium states. This behavior has been clearly observed in out-of-equilibrium experiments on the overstretching transition \cite{WhitelamBJ2008,FuNAR2010,BiancoBJ2011}. The hysteresis curves are consequence of force ramps applied on a time scale shorter than the DNA relaxation time. Using waiting times of the order of 1 min the equilibrium state is recovered and no hysteresis is observed \cite{leger1999}.
Figure~\ref{ist} shows how our model mimics this behavior. In fact, the left panel of the figure presents a typical hysteresis curve obtained by decreasing the simulation time, i.e., not allowing the system to reach equilibrium. For a simulation time of $t_s = 10^6$~t.u., we already have a different path for increasing forces than for decreasing ones (indicated in the figures by the black arrows). For a shorter simulation time $t_s = 3 \times 10^5$~t.u., the hysteresis area is bigger, while for $t_s = 5 \times 10^6$~t.u. (squares in the figure), no hysteresis is observed. This behavior expresses a typical dynamical property of the model not visible in a static equilibrium analysis.

Finally, we discuss the effect of the $k_l$ parameter on the transition. The right panel of Fig.~\ref{ist} shows how this parameter is responsible in the model for the force value where the transition occurs. This value can be related to those parameters whose magnitude affects the overstretching transition: the value of $p$H \cite{WilliamsBJ2001} or the salt concentration \cite{WennerBJ2002,Fu2009}.

{\it Discussion}.
The picture of the DNA overstretching transition is getting richer and richer as the number of experiments increases. However, as we stated in the Introduction, the microscopic nature of the transition is still controversial.  An interesting approach to the problem is to study single-stranded DNA \cite{ke2007}. Although the elongation features in that case could give some insight in explaining the rich elastic properties of DNA, the presence of two plateaus in single-stranded DNA found in \cite{ke2007} shows that the elasticity of DNA is much more complex than previously thought.
It seems difficult to put together the features of a double-stranded DNA with those of single-stranded DNA, where the force-elongation characteristics depend on the nature of the specific piece of DNA object of elongation.

Possibly related to this, other experimental works suggest that DNA melting could be responsible for the DNA overstretching features \cite{WilliamsBJ2001,WennerBJ2002,MetzlerPRL2008,WilliamsPNAS2009,vanMameremPNAS2009}. Although strong evidence in that direction appears reasonable, many doubt still remain in this respect \cite{DanilowiczPNAS2009,FuNAR2010,BiancoBJ2011,ZhangPNAS2012}.

The model presented here does not consider any microscopic aspects of the transition but rather adopts, in the GL spirit, a mesoscopic point of view, valid, in principle, for any of the above descriptions. We consider the transition between two states (the B-DNA  and the overstretched state) using a stochastic equation of motion with an asymmetric double-well potential in the presence of one free parameter connected to biophysical properties of the DNA chain and/or of the surrounding solvent. The model is able to reproduce the equilibrium (over)stretching curve with good precision. Moreover, the approach used here allows us to study the {\it dynamics} of the transition. It can be used to carry out realistic calculations in DNA dynamics, as in translocation under strong forces caused by molecular and nanotechnological motors, and represents a valid tool to perform out-of-equilibrium modeling. In this sense the model can be expanded in order to include other variables able to describe more specific microscopic pictures, such as, for example, the twist-stretch coupling previously discussed, or melting properties.

\vspace{0.5cm}
\emph{Acknowledgments.} This work has been supported by the Spanish DGICYT Projects No. FIS2008-01240 and No. FIS2011-25167, co-financed by Fondo Europeo de Desarrollo Regional (FEDER) funds. Financial support from European Science Foundation Research Network "Exploring the Physics of Small Devices" is acknowledged. We also want to thank Professor Alessandro Pelizzola for useful conversations.
\vspace{0.5cm}

\end{document}